\documentclass[12pt]{article}
\usepackage{graphicx}
\usepackage{epsfig}
\usepackage{psfrag}
\usepackage{natbib}
\usepackage{txfonts}
\usepackage{bm}
\def\dsize{\displaystyle}

\begin{document}

\begin{center} {\Large\bf
The parametric resonance   in the $\alpha\omega$-dynamo models}
\end{center}

\begin{center} {\large 
Maxim Reshetnyak}
\end{center}

\begin{center} {\small\it 
Institute of the Physics of the Earth RAS\\ B.Gruzinskaya 10, Moscow, 123995,  Russia \\
{\it email}: {\tt m.reshetnyak@gmail.com} }
\end{center}



\begin{abstract}{\small \noindent 
It is shown, that retardation in the $\alpha$-quenching in the Parker's dynamo model leads to parametric resonance. This result is observed in the numerical simulations and can be reproduced in the simple analytic model. The other interesting effect in the model with retardation is a appearance of the long-term processes with the period much larger than the typical time of retardation.
\vskip 0.5cm
 \noindent {\bf Keywords:} {dynamo, instabilities, magnetic fields, turbulence.}}
\end{abstract}

              
\section{Introduction}

 The main idea of the dynamo theory, which is believed can explain existence of the 
  magnetic fields observed in cosmos,  
    is that kinetic energy of the conductive motions is transformed into the energy of the magnetic field.    
     Magnetic field generation is the threshold phenomenon: it starts when magnetic Reynolds number $\rm R_m$ reaches its critical value $\rm R_m^{\rm cr}$. After that magnetic field grows exponentially up to the moment, when it  already can feed back on the flow. Description of transition from the linear regime when influence of the magnetic field onto the flow is negligible to the nonlinear regime, when magnetic energy can exceed the kinetic energy orders of magnitude (like it is in the planetary cores) is a subject of the modern researches  \cite{BS2005}. 
 
 As a result, even after quenching the saturated velocity field is still large enough, so that   $\rm R_m\gg R_m^{\rm cr}$. Moreover, velocity field taken from the nonlinear problem (when the exponential growth of the magnetic field stopped) can still generate exponentially growing magnetic field providing that the feed back of the magnetic field on the flow is omitted (kinematic dynamo regime) \cite{CT2009, T2008, TB2008,  SSCH2009, HR2010}. 
 It appears,  that  problem of  stability  of the full dynamo equations including induction equation, the Navier-Stokes equation with the Lorentz force differs  from the stability problem of the single induction equation with the given saturated velocity field taken from the full dynamo solution: stability of the first problem does not provide stability of the second one. Moreover,  some regimes
 close to the Case 1 from the geodynamo benchmark  \cite{Ch2001} are stable in contrast to the solutions with the periodical boundary conditions in space and influence of the boundary conditions can be important \cite{T2008,   SSCH2009}. 

One of the simple explanations of this phenomenon (at least for some regimes) was offered in \cite{Resh2010}. Using Parker's dynamo model for the thin disk it was shown, that  the $\alpha$-effect, taken from the nonlinear oscillating saturated problem, can still generate exponentially growing magnetic field. The origin of this effect is closely related to  the parametric resonance. Here we develop these ideas and show how the phase shift in the $\alpha$-quenching can effect on the behaviour of the generated magnetic field.
 
\section{Parker's dynamo}
One of the simplest  dynamo models used in the solar and galactic applications 
 is the Parker's one-dimensional model \cite{Parker1971} (see its development for the galactic dynamo in 
\cite{RSS1988}):
       \begin{equation}\begin{array}{l}\dsize
    {\partial A\over  \partial  t} =\alpha B
 + A'',   \qquad  
{\partial B\over  \partial  t} =-{\cal D} A'+ B'', 
\end{array}\label{sys11}
\end{equation}
where   $A$ and $B$ are   azimuthal components of the vector potential  
and magnetic field, $\alpha(z)$ is a kinetic helicity,  
${\cal D}$ is a dynamo number, which is a product of the amplitudes of the $\alpha$- and $\omega$-effects and  primes denote derivatives  with respect to a coordinate. For the Galaxy the only one left coordinate is 
 cylindrical polar coordinate $z$. For the thin shells, which we will have in mind in this paper, is 
 a latitude  $\vartheta$.
  Equation (\ref{sys11}) is solved in the interval $-90^\circ \le  \vartheta\le 90^\circ$ with the boundary conditions 
$B=0$ and $A'=0$ at $\vartheta=\pm 90^\circ$.
 System  (\ref{sys11}) has growing solution, when $|{\cal D}|>|{\cal D}^{\rm cr}|$.  
  Putting nonlinearity of the form
 \begin{equation}\begin{array}{l}\dsize
 \dsize \alpha(\vartheta)={\alpha_0(\vartheta)\over 1+E_m} 
 \end{array}\label{non}
\end{equation} 
in  (\ref{sys11}), where  $\dsize E_m=( B^2+{ A'}^2)/2$ is a magnetic energy, gives  quasi-stationary solutions for the positive $\cal D$, see about various forms of nonlinearities in \cite{Beck}. The property of the nonlinear solution is mostly predetermined 
  by the form of its  first  eigenfunction.

The main result of \cite{Resh2010} was, that 
 simultaneous solution of equations for $A,\, B,\, \alpha$ (\ref{sys11}, \ref{non}) and 
similar equations for the new magnetic field $(\widehat{A},\, \widehat{B})$:
       \begin{equation}\begin{array}{l}
       \dsize
    {\partial \widehat{A}\over  \partial  t} =\alpha \widehat{B}
 + \widehat{A}'',  \qquad   
{\partial \widehat{B}\over  \partial  t} =-{\cal D} \widehat{A}'+ \widehat{B}''. 
\end{array}\label{sys22}
\end{equation}
with the same $\alpha$ can lead to the exponentially growing $(\widehat{A},\, \widehat{B})$, when the 
field $(A,\, B)$ is already saturated. It happens when  $(A,\, B)$ oscillates,  and 
initial conditions for $(A,\, B)$ and  $(\widehat{A},\, \widehat{B})$ are slightly different. 
 This effect is very similar to what was observed in the more sophisticated models \cite{CT2009},  \cite{TB2008}, \cite{HR2010}.  Below we consider this effect in more details.

 \section{Parker's dynamo with retardation}
\begin{figure}[th!]
 \psfrag{Em}{ $\overline{B^2}$}
  \psfrag{t}{ $\tau$}
\hskip 3cm 
  \includegraphics[width=7cm]{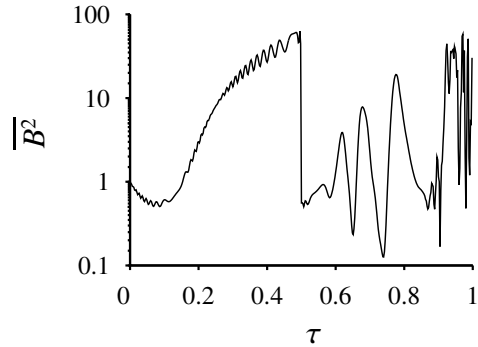}
 \caption{
Dependence of the mean over the volume and time squared magnetic field   
  $B^2$ on retardation $\tau$ in $\alpha$-quenching. 
 $\overline{B^2}$ is normalized in such a way, that $\overline{B^2}\big{|}_{\tau=0}=1$. $\tau$ is measured in units  of $T_0$.
  } \label{fig1}
\end{figure}

\begin{figure}[th!]
\vskip -.4cm
\psfrag{t}{ $t$}
\psfrag{Em}{ $E_m$}
\begin{minipage}[t]{.35\linewidth}
 \psfrag{GH}{ a}\vskip 0cm  \hskip -0.5cm \epsfig{figure=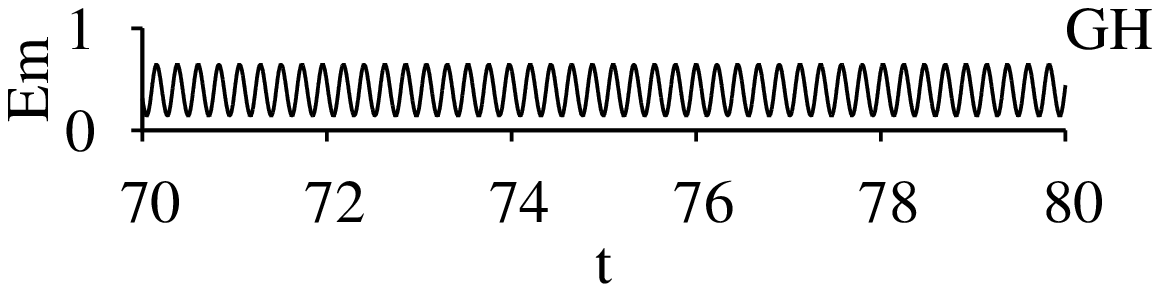,width=7cm}
\end{minipage}\hfill
\begin{minipage}[t]{.35\linewidth}
 \psfrag{GH}{ b}\vskip 0cm  \hskip -2.5cm   \epsfig{figure=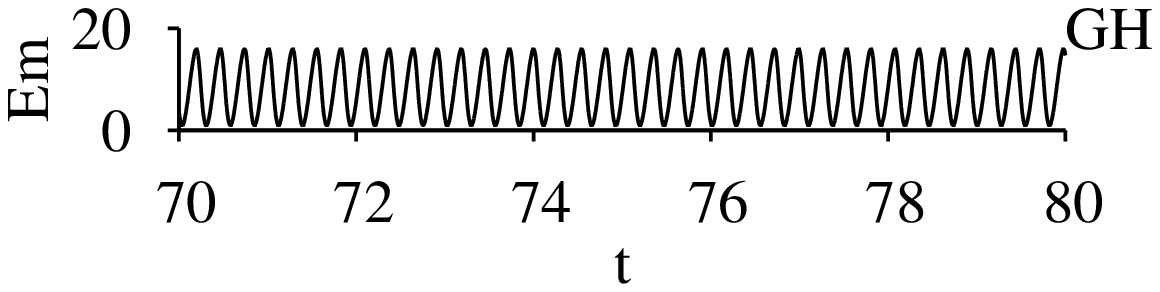,width=7cm}
\end{minipage}\hfill
\begin{minipage}[t]{.35\linewidth}
 \psfrag{GH}{ c}\vskip 0cm  \hskip -0.5cm   \epsfig{figure=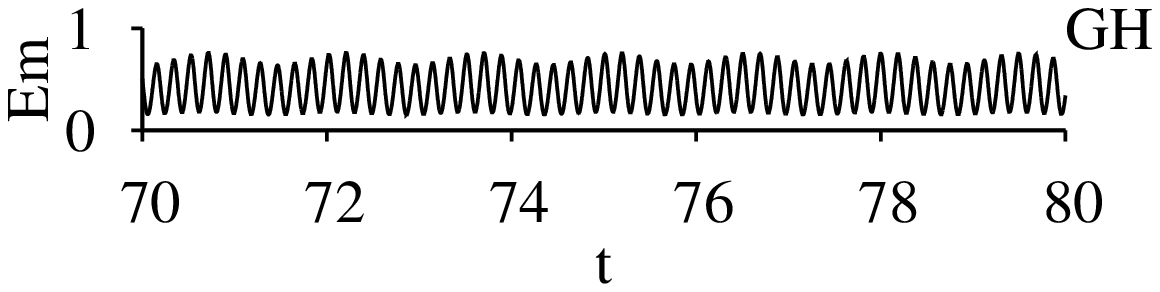,width=7cm}
 \end{minipage}\hfill
\begin{minipage}[t]{.35\linewidth}
  \psfrag{GH}{ d}\vskip 0cm \hskip -2.5cm   \epsfig{figure=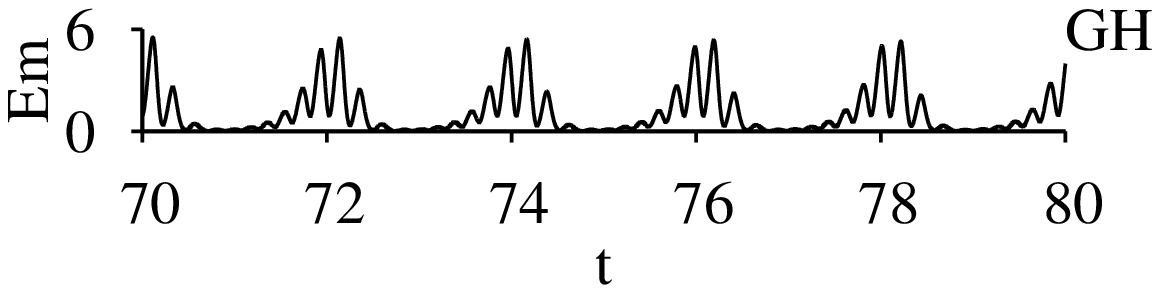,width=7cm}
  \end{minipage}\hfill
\begin{minipage}[t]{.35\linewidth}
 \psfrag{GH}{e}\vskip 0cm  \hskip -0.75cm   \epsfig{figure=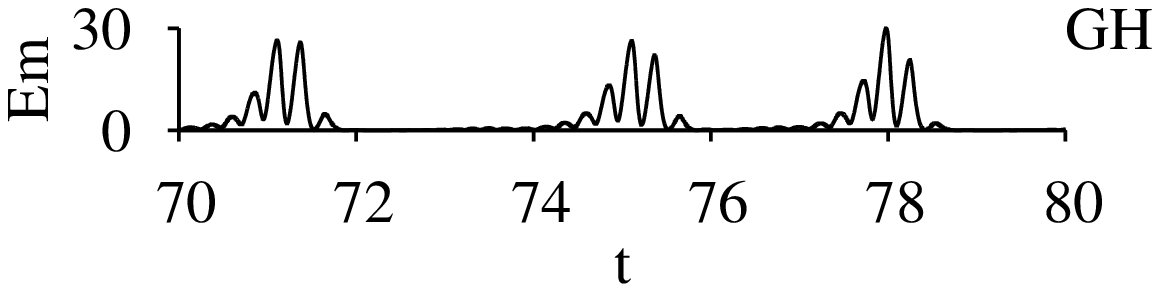,width=7cm}
 \end{minipage}\hfill
\vskip 0.0cm
 \caption{
Evolution of the mean over the volume magnetic energy for the different retardation: 
 $\tau=0$ (a),  $\tau=0.33$ (b),
  $\tau=0.57$ (c),  $\tau=0.67$ (d), $\tau=0.89$ (e). 
 } \label{fig2}
\end{figure}

 Here we return to the system  (\ref{sys11}). 
 So as  $\alpha$ is a function of the magnetic field which oscillates, in the general case we are in a position to expect 
  appearance of the parametric resonance for $(A,\, B)$, as well. Now we consider the more general form of the nonlinearity with  retardation 
      $\tau$:  
        \begin{equation}\begin{array}{l}\dsize
 \dsize \alpha(\vartheta,\, t,\,\tau)={\alpha_0(\vartheta)\over 1+B^2(\vartheta,\, t-\tau)},
\end{array}\label{non2}
\end{equation} and find how $\tau$ 
 effects on the  magnetic field  generation.  
  Results for  $\alpha_0=\sin(2\vartheta)$ and ${\cal D}=300$ shown in 
 Fig.~\ref{fig1} are quite unexpected: in spite of the fact, that  $\alpha$ 
 depends  on the squared magnetic field, the mean magnetic energy is not symmetric on 
   $\tau=0.5$ ($\tau$ is in units of process's half period  $T_0\approx 0.45$).    
 After some decrease of the magnetic field amplitude for $0<\tau<\tau_{min}=0.17$
 magnetic field starts to increase being periodical,
  see Fig.~\ref{fig2}.
   The sharp decrease at   
    $\tau=\tau_{br}\approx 0.5$  (see Fig.~\ref{fig2}c) takes place up to the 
      values at  $\tau_{min}$, accompanied by the 
       long-term modulation with the period 8 times larger than the period of the 
 original oscillation of $E_m$ $T_0$ at         
        $\tau=0$. The amplitude of this oscillation increases with increase of 
$\tau$, see Fig.~\ref{fig2}d-e. The amplitude of the new oscillation starts to change at 
   $\tau=0.89$, resembling beating, see Fig~\ref{fig2}e.

\begin{figure}[h!]\centering{
\vskip -0.2cm
\psfrag{t}{$t$}
\begin{minipage}[!h]{.45\linewidth}
\psfrag{Br}{ $B_r$}\psfrag{GH}{a}\epsfysize=1.8cm  \hskip -0.4cm \epsfbox{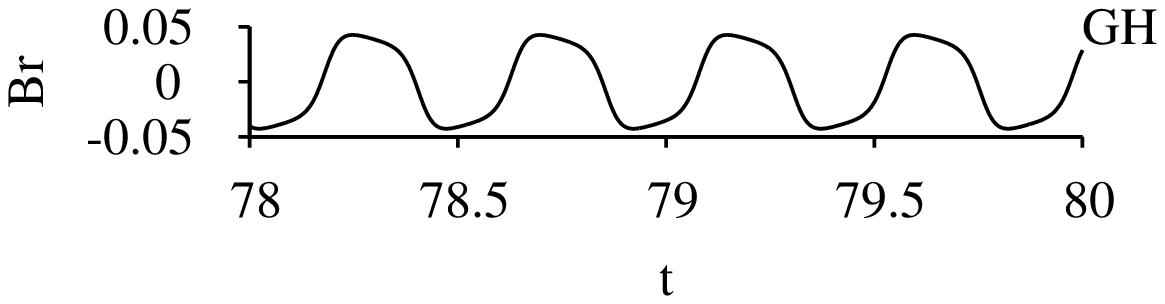}
\end{minipage}\hfill
\begin{minipage}[!h]{.45\linewidth}
\psfrag{B}{ $B$}\psfrag{GH}{ b}\epsfysize=1.8cm  \hskip -1cm \epsfbox{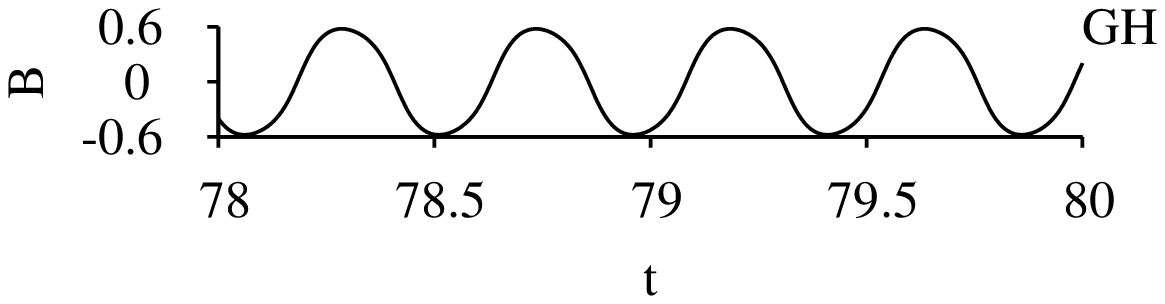}
\end{minipage}\hfill

\begin{minipage}[!h]{.45\linewidth}
\psfrag{Br}{ $B_r$}\psfrag{GH}{ c}\epsfysize=1.8cm  \hskip -0.4cm \epsfbox{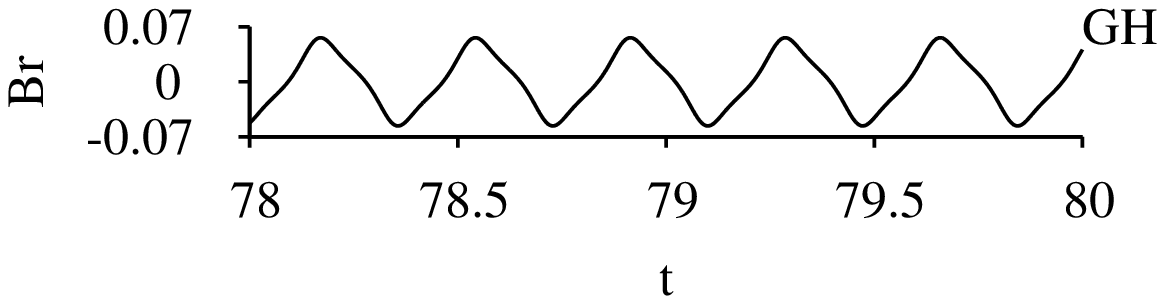}
\end{minipage}\hfill
\begin{minipage}[!h]{.45\linewidth}
\psfrag{B}{ $B$}\psfrag{GH}{ d}\epsfysize=1.8cm  \hskip -1cm \epsfbox{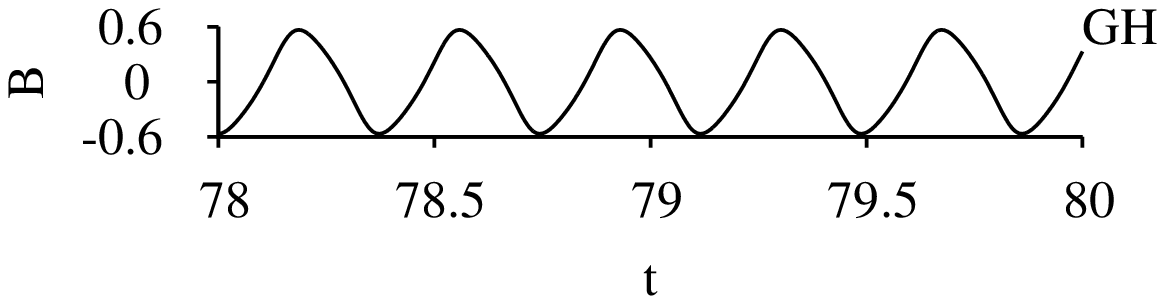}
\end{minipage}\hfill

\begin{minipage}[!h]{.45\linewidth}
\psfrag{Br}{ $B_r$}\psfrag{GH}{ e}\epsfysize=1.8cm  \hskip -0cm \epsfbox{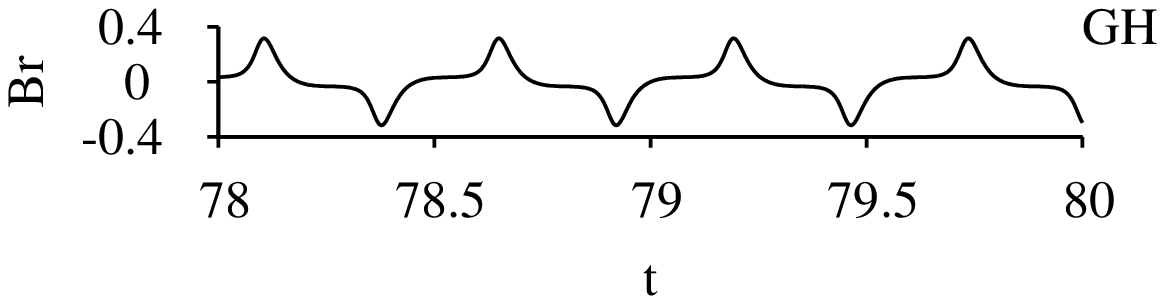}
\end{minipage}\hfill
\begin{minipage}[!h]{.45\linewidth}
\psfrag{B}{ $B$}\psfrag{GH}{ f}\epsfysize=1.8cm  \hskip -1cm \epsfbox{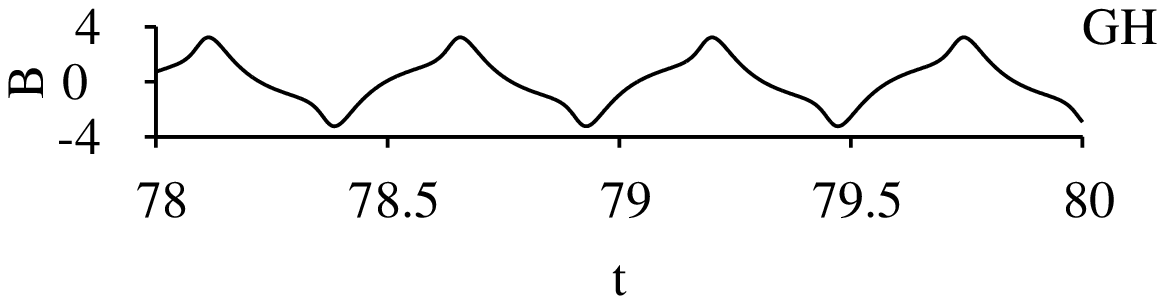}
\end{minipage}\hfill

\begin{minipage}[!h]{.45\linewidth}
\psfrag{Br}{ $B_r$}\psfrag{GH}{ g}\epsfysize=1.8cm  \hskip -0cm \epsfbox{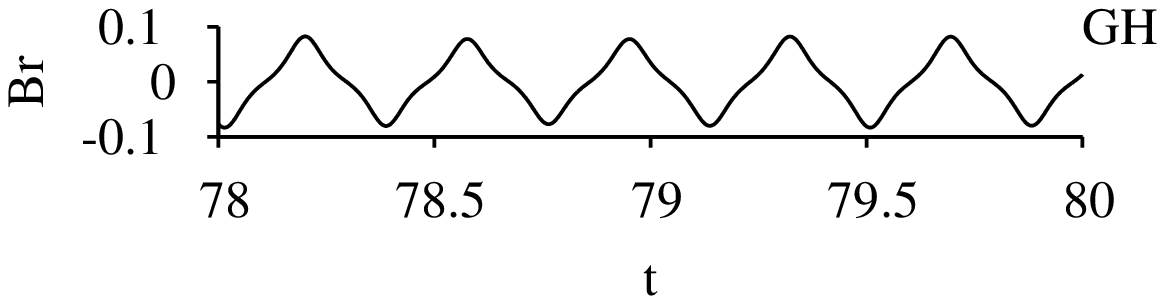}
\end{minipage}\hfill
\begin{minipage}[!h]{.45\linewidth}
\psfrag{B}{ $B$}\psfrag{GH}{ h}\epsfysize=1.8cm  \hskip -1cm \epsfbox{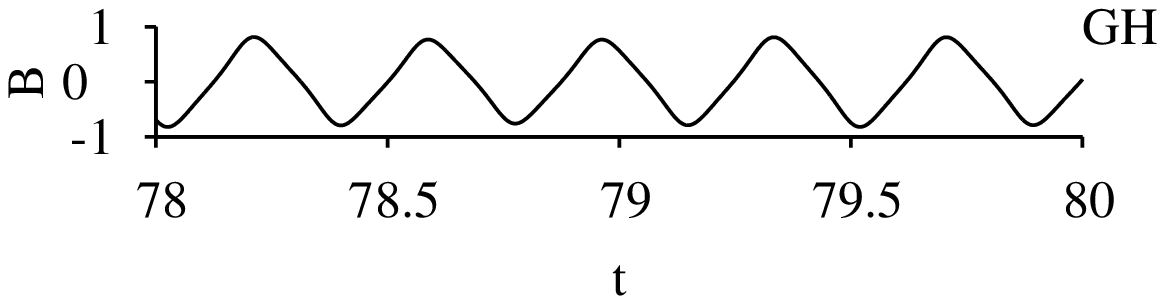}
\end{minipage}\hfill

\begin{minipage}[!h]{.45\linewidth}
\psfrag{Br}{ $B_r$}\psfrag{GH}{ i}\epsfysize=1.8cm  \hskip -0cm \epsfbox{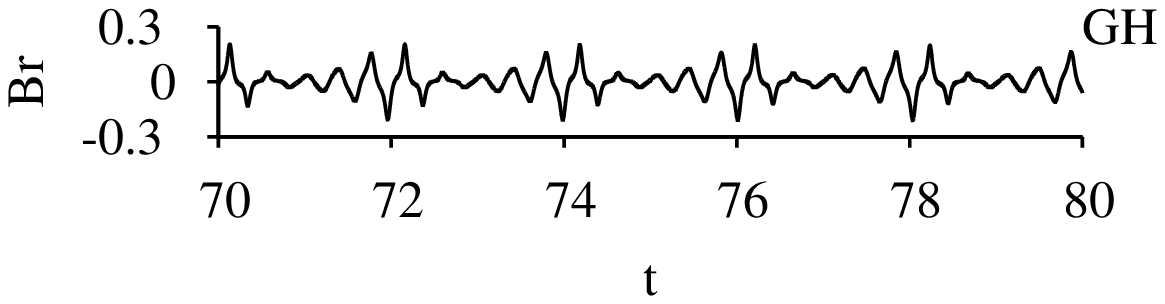}
\end{minipage}\hfill
\begin{minipage}[!h]{.45\linewidth}
\psfrag{B}{ $B$}\psfrag{GH}{ j}\epsfysize=1.8cm  \hskip -1cm \epsfbox{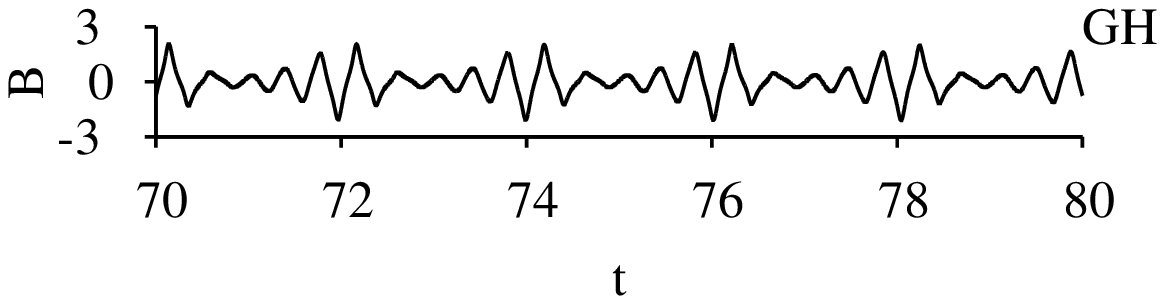}
\end{minipage}\hfill

\begin{minipage}[!h]{.45\linewidth}
\psfrag{Br}{ $B_r$}\psfrag{GH}{ k}\epsfysize=1.8cm  \hskip -0cm \epsfbox{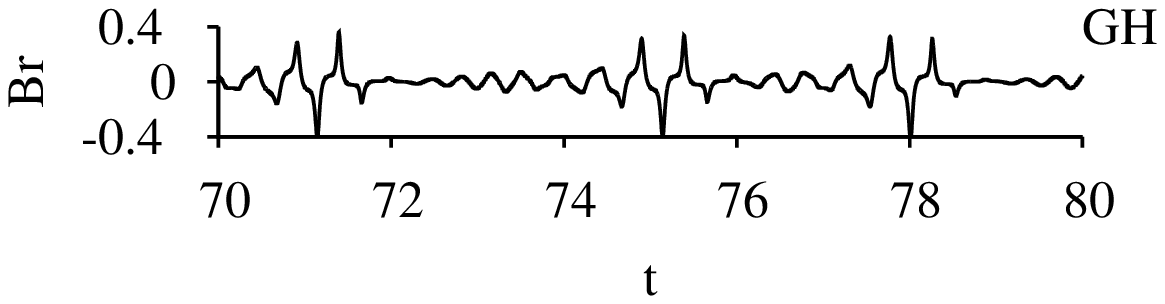}
\end{minipage}\hfill
\begin{minipage}[!h]{.45\linewidth}
\psfrag{B}{ $B$}\psfrag{GH}{ l}\epsfysize=1.8cm  \hskip -1cm \epsfbox{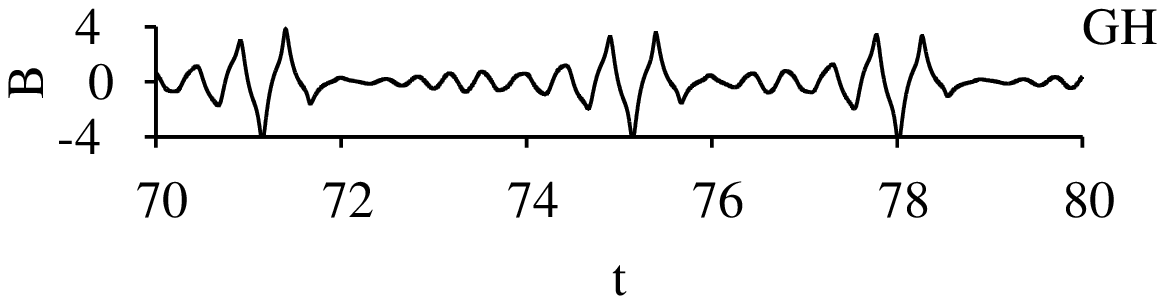}
\end{minipage}\hfill
}\vskip 0cm
\caption{
Evolution of the mean over the half of the  volume $B_r$ (left column) and  $B$ (right column) for  
  $\tau=0$ -- a, b; 0.1 -- c,d; 0.33 -- e,f; 0.57 -- g,h; 0.67 -- i,j; 0.89 -- k,l.  
} \label{fig3}
\end{figure}

Increase of  $\tau$ up to  0.1 leads to the change of the form of the radial $B_r=-A'_\vartheta$ and azimuthal 
$B$ components  from sinusoidal to the 
saw-shaped form, see Fig.~\ref{fig3}, accompanied with delay of $B_r$ from $B$. For the larger 
  $\tau$ the phase shift $\varphi$ decreases. 
 Increase of $\tau$ to  0.33 leads to the growth of the magnetic field and appearance of 
  the pike-like extremums.  This region of $\tau$ corresponds to the parametric resonance.    
\pagestyle{empty}

  The further increase of $\tau$ leads to the sharp decrease of the magnetic field amplitude, see Figs.
  ~\ref{fig1}, \ref{fig2}, \ref{fig3}.
  The form of the curve is close to that one for    $\tau=0.1$ with one exception: 
 the new large period appears.  Note, that in Yoshimura (1978a, 1978b) 
  such long-term modulation was  riched at $\tau>T_0$, see review  of some other 
   sources of the long-term variations in the solar dynamo in  \cite{Tob2002}. The different behaviour of   ${E_m}$ 
 for  $0<\tau<0.5$ and  $0.5<\tau<1$ may be explained by the 
 accumulation of the time shift between the magnetic field and $\alpha$. 
This suggestion is supported by the fact that for $\tau>1$ behaviour of 
  ${E_m}$ is very close to that one for $\tau>0.7$.
\pagestyle{empty}
\begin{figure}[th!]
 \psfrag{x}{ $t$}  \psfrag{y}{ $\vartheta$}
 \psfrag{GH}{a}
 \begin{minipage}[!h]{.45\linewidth}
 \vskip 0cm  \hskip 0cm \epsfig{figure=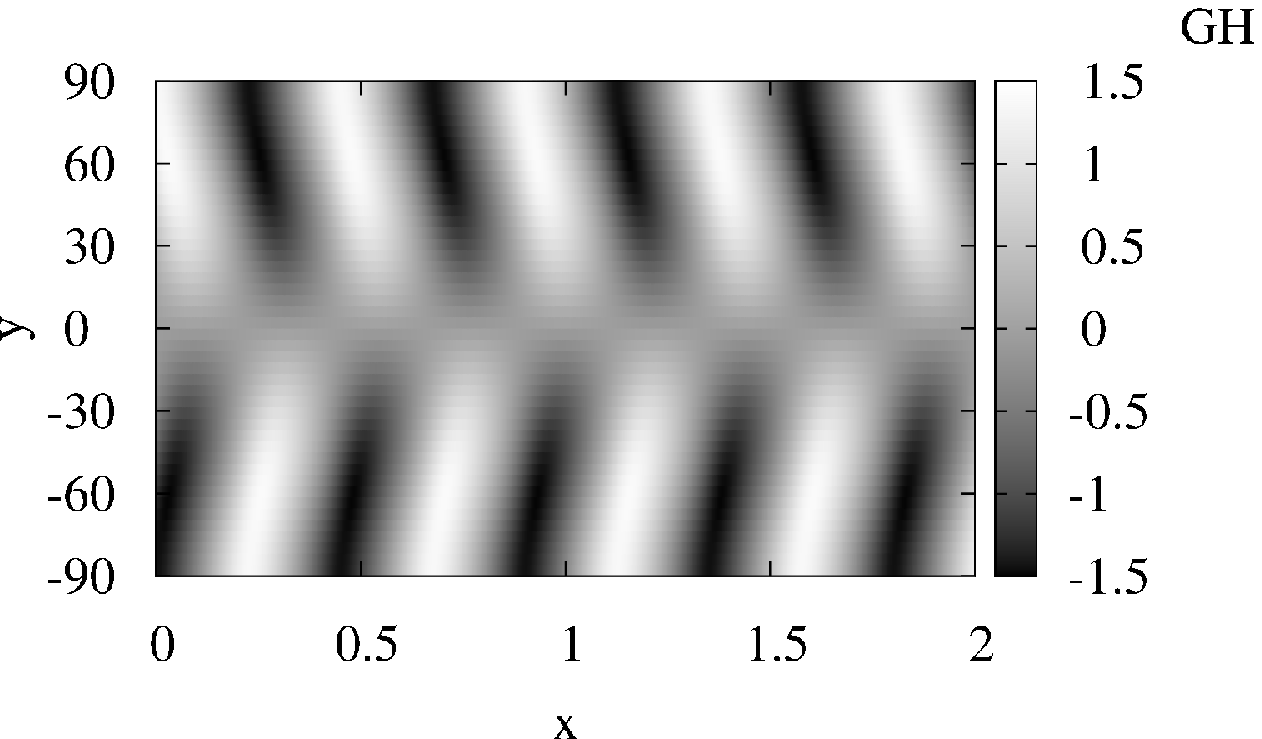,width=8cm}
 \end{minipage}\hfill
 \begin{minipage}[!h]{.45\linewidth}
 \psfrag{GH}{b}\vskip -0cm  \hskip 1cm \epsfig{figure=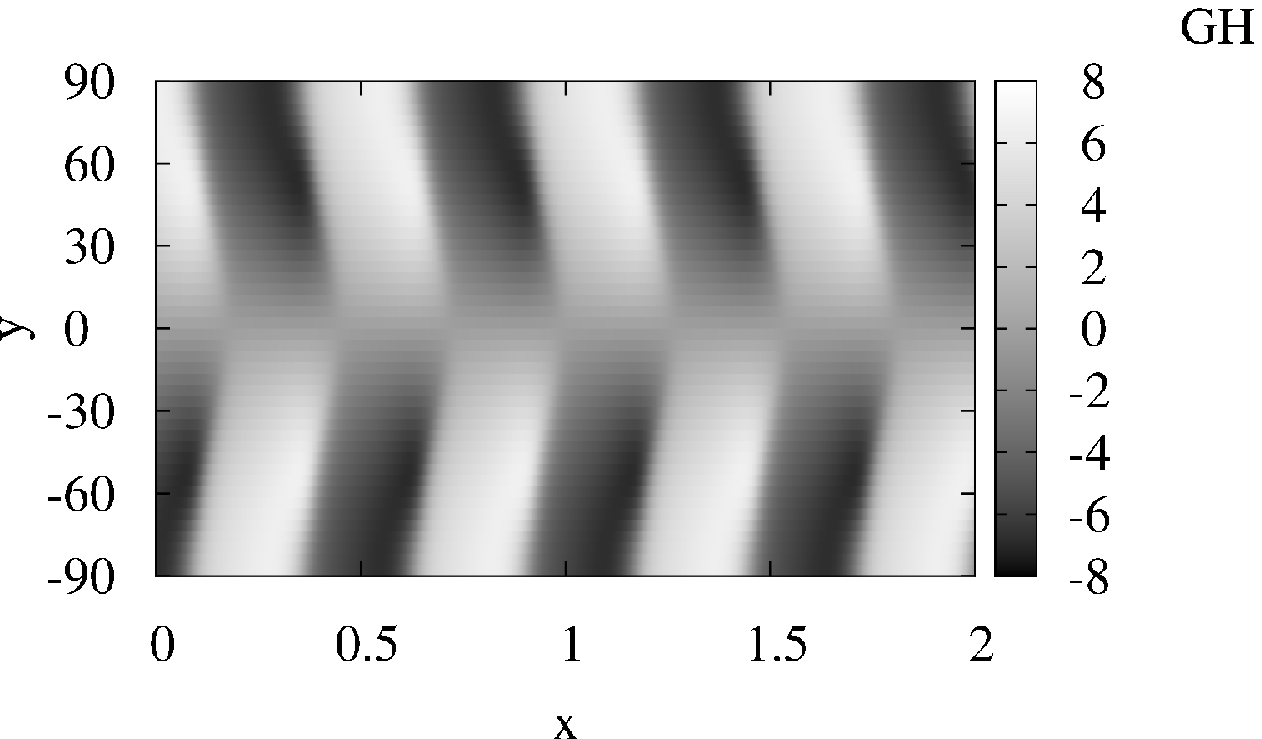,width=8cm}
 \end{minipage}\hfill
 \begin{minipage}[!h]{.45\linewidth}
 \psfrag{GH}{c}\vskip -1cm  \hskip -0.1cm \epsfig{figure=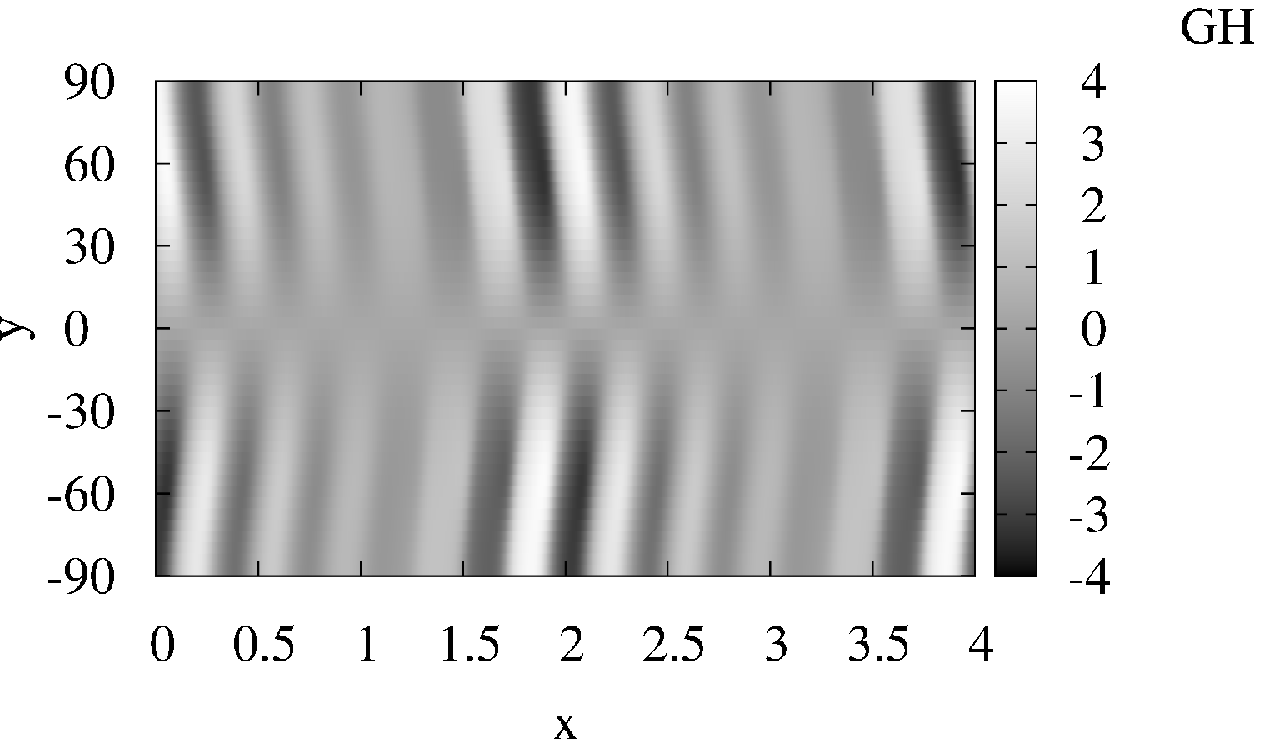,width=8cm}
 \end{minipage}\hfill
 \caption{ The butterfly diagram  $B(t,\,\theta)$:  
  $\tau=0$ -- a;  $0.33$ -- b and $0.67$ -- c.   
  } \label{fig4}
\end{figure}
 
Up to the moment we did not consider how the magnetic field depends on 
  $\vartheta$. 
 The dipole solution is a wave propagating from the poles to the equator
  with maximum at the middle latitudes for   $\tau=0$, see Fig.~\ref{fig4}.  
   Increase of $\tau$  leads to the stripe-like solution, i.e. more contrast changes of the sign of the fields and increase of the magnetic field at the poles Fig.~\ref{fig4}b. The long term variations are well resolved in Fig.~\ref{fig4}c.

\section{Parametric resonance}   
To consider how parametric resonance appears we follow  \cite{Resh2010} and present solution for (A,\, B) in the form of the waves: 
$B=b\sin(t)$, ${A}=\sin(t+\varphi+\theta)$, $ {B}= \sin(t+\theta)$, and 
 $\dsize\alpha={1\over 1+B^2(t-\tau)}$ we get how generation depends on $\tau$. 
  Then putting it in (\ref{sys11}) we get two equations. 
   Equation for ${B}$ does not include  $\tau$, so we consider only production of ${A}^2$. Then 
  $\delta {A}(\varphi,\,\theta)=\alpha_0\int\limits_0^{2\pi}{\dsize {B}{A}\over\dsize 1+B^2}\, dt$. 
  If 
  $\dsize|\Pi|\gg 1$, where $\dsize\Pi={\delta {A}(\varphi,\,\theta)\over \delta {A}(\varphi,\,0)}$,  then $(A,\, {B})$ is  unstable.
  
   The integral for  $\delta A$ gives:
       \begin{equation}\begin{array}{l}
       \dsize 
\delta {A}(\varphi,\,\tau)=h_1  
  +
 h_2  {\rm tg}(\varphi),
\qquad
\dsize 
h_1\approx 1-0.3\cos(\tau)^2, \quad
\qquad
h_2\approx 0.8\sin(2\tau).
\end{array}\label{sys44aa1}
\end{equation} 

\begin{figure}[th!]
\vskip -0cm
  \psfrag{t}{ $\tau$}
    \psfrag{A}{ $\Pi$}
\hskip 3cm \epsfig{figure=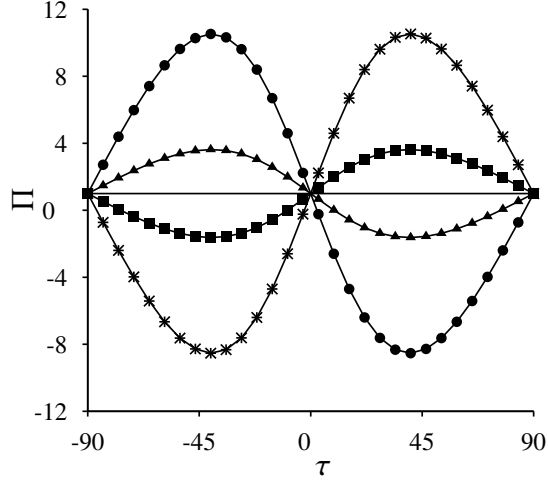,width=8cm}
 \caption{Dependence of $\Pi(\tau)$ for different $\varphi$: 
0 -- solid line, 
$-85^\circ$ -- circles, 
$-70^\circ$ -- triangles, 
$70^\circ$ -- squares, 
$85^\circ$ -- stars. The straight line corresponds to $\Pi=1$.
} \label{fig5}
\end{figure}

Then  
       \begin{equation}\begin{array}{l}
       \dsize 
\Pi=1+{0.8\sin(2\tau)\over 1-0.3\cos(\tau)^2}{\rm tg}(\varphi) 
\end{array}\label{sys44aa1}
\end{equation}  
and for  $\dsize \varphi\to \pm {\pi\over 2}$ (what corresponds to  the phase shift between the components 
 $\varphi_{B_r\,B}\approx 0$)  
and  $\tau\ne 0$ $|\Pi|$ grows, and parametric resonance appears, see Fig.~\ref{fig5}. 
Note, that this analysis explains small decrease of $B^2$ at small $\tau$ (see Fig.~\ref{fig1}) which corresponds to the positive values of $\varphi$ in  Fig.~\ref{fig5}.
\pagestyle{empty}

\section{Conclusions}
Here we considered the only one form of the nonlinearity for the fixed value of the dynamo number $\cal D$ as a function of the time lag  $\tau$ in the $\alpha$-quenching. However even  this simple model demonstrates variety of effects: change of the form of the poloidal and toroidal fields, regions of the weaker and stronger fields, appearance of the long-term variations. The quite natural choice of $\tau$ can lead to the sharp increase of the  
 magnetic field amplitude concerned with the parametric resonance. This explanation does not contradict to the simple linear analysis presented above. It is very tempting to correspond appeared 
 in simulations the  long-term periodicities with that ones of the solar activity larger 
  than the main 22 years period. The difficulty is to justify the choice of the particular $\tau$ which should be obtained from the solution of the more sophisticated nonlinear model.

\vskip 0.5cm
\noindent 
I thank   D.Sokoloff for discussions.

\label{lastpage}


\begin{thebibliography}{99}
\bibitem[Beck et al.  (1996)]{Beck} {Beck, R., Brandenburg, A., Moss D., Shukurov,  A. \&
  Sokoloff, D.}  1996, \textit{ARA$\&$A}, 34, 155


\bibitem[Brandenburg \& Subramanian (2005)]{BS2005} {Brandenburg, A. \& Subramanian, K.} 2005,
\textit{Phys. Rep.}, 41, 1

\bibitem[Cattaneo \& Tobias (2009)]{CT2009} 
{Cattaneo, F. \& Tobias, S.M.} 2009, \textit{J. Fluid Mech.}, 621, 205, arXiv:0809.1801

\bibitem[Christensen et al. (2001)]{Ch2001} 
{ Christensen, U.R., Aubert, J., Cardin, P., Dormy, E., Gibbons,  
S., Glatzmaier, G.A., Grote, E., Honkura, Y., Jones, C., Kono, M.,
Matsushima, M., Sakuraba, A., Takahashi, F., Tilgner, A., Wicht,    
J. \& Zhang, K.} 2001,   \textit{Phys. Earth Planet. Inter.}, 128, 25


\bibitem[Hejda \& Reshetnyak (2010)]{HR2010} {Hejda, P \& Reshetnyak, M.} 2010, 
\textit{Accepted to Geophys. Astrophys. Fluid Dynam.},  104, 
arXiv:1005.1557




\bibitem[Parker (1971)]{Parker1971} Parker, E.N. 
 1971, \textit{Astrophys. J.}, 163,  255 


\bibitem[Reshetnyak (2010)]{Resh2010} {Reshetnyak, M.} 2010, 
\textit{ Mon. Not. R. Astron. Soc.}  405, L90, 
                         arXiv:1001.4234 

\bibitem[Ruzmaikin,  Shukurov \&  Sokoloff (1988)]{RSS1988} 
{Ruzmaikin, A. A. Shukurov, A. M. \& Sokoloff, D. D.} 1988, Magnetic Fields in Galaxies (Kluwer Academic Publishers,  Dordrecht), 280

\bibitem[Schrinner, Schmitt, Cameron (2010)]{SSCH2009} {Schrinner, M., Schmitt, D., Cameron, R. \& Hoyng, P.} 2010, \textit{Geophys. J. Int.} 182, 675, arXiv: 0909.2181.


\bibitem[Tilgner (2008)]{T2008} {Tilgner, A.} 2008, \textit{Phys. Rev. Lett.}, 100, 128501

\bibitem[Tilgner \&  Brandenburg (2008)]{TB2008} {Tilgner, A. \&  Brandenburg, A.} 2008,
\textit{ Mon. Not. R. Astron. Soc.}, 391, 1477,  arXiv:0808.2141 

\bibitem[Tobias (2002)]{Tob2002}  {Tobias, S.M.}. 2002,
\textit{Astron. Nachr.}, 323, 417


\bibitem[Yoshimura  (1978a)]{Yosh1978a}  {Yoshimura, H.}. 1978a,
\textit{ApJ}, 221, 1088

\bibitem[Yoshimura  (1978b)]{Yosh1978b}  {Yoshimura, H.}. 1978b,
\textit{ApJ}, 226, 706


\end{thebibliography}
\end{document}